\documentclass[aps,prl,numerical,reprint,onecolumn,prl,showpacs,amsfonts,amssymb,amsmath,mathbbol,letterpaper,groupedaddress,nobibnotes,footinbib,floatfix]{revtex4-1}

\usepackage{amsmath}    % need for subequations
\usepackage{graphics}   % need for figures
\usepackage{color}      % use if color is used in text
\usepackage{epsfig}
\usepackage{epstopdf} % need for mac
\usepackage{bm}        % for math
\usepackage{verbatim}   % for math
\usepackage{xspace}

\usepackage{graphicx}
\usepackage{multirow}
\usepackage{rotating}
\usepackage{amsfonts}
\usepackage{color}
\usepackage[T1]{fontenc}
\usepackage{etoolbox}
\usepackage{lineno}
\usepackage[colorlinks=true, linkcolor=blue,citecolor=blue,urlcolor=blue]{hyperref}
\newcommand{\overbar}[1]{\mkern 1.5mu\overline{\mkern-1.5mu#1\mkern-1.5mu}\mkern 1.5mu}

\begin{document}
\draft \narrowtext \sloppy
\title{\Large Oxygen-enabled control of Dzyaloshinskii-Moriya Interaction in ultra-thin magnetic films}

\author{ \large Abderrezak\,Belabbes$^{1}$}
\email [ \normalsize  Electronic \ address: ]{abderrezak.belabbes@kaust.edu.sa}
\author{\large Gustav\,Bihlmayer$^{2}$}
\email[\normalsize  Electronic \ address: ]{g.bihlmayer@fz-juelich.de}
\author{\large Stefan\,Bl\"ugel$^{2}$}
\email[\normalsize  Electronic \ address: ]{s.bluegel@fz-juelich.de}
\author{\large Aur\'elien\,Manchon$^{1}$}
\email[\normalsize  Electronic \ address: ]{aurelien.manchon@kaust.edu.sa}

\affiliation{$^1$
\mbox{\fontsize{10.5}{10}\selectfont   King Abdullah University of Science \& Technology (KAUST), Thuwal 23955-6900, Saudi Arabia}}
\affiliation{$^2$
\mbox{\fontsize{10.5}{10}\selectfont  Peter Gr\"unberg Institut \& Institute for Advanced Simulation, Forschungszentrum J\"ulich \& JARA }
\mbox{\fontsize{10.5}{10} \selectfont D-52425 J\"ulich, Germany}}

\begin{abstract} 

\vspace{1cm}
\hspace{6cm}{\Large Abstract} \\

\fontsize{10.5}{10}\selectfont  The search for chiral magnetic textures in systems lacking spatial inversion symmetry has attracted a massive amount of interest in the recent years with the real space observation of novel exotic magnetic phases such as skyrmions lattices, but also domain walls and spin spirals with a defined chirality. The electrical control of these textures offers thrilling perspectives in terms of fast and robust ultrahigh density data manipulation. A powerful ingredient commonly used to stabilize chiral magnetic states is the so$-$called Dzyaloshinskii$-$Moriya interaction (DMI) arising from spin$-$orbit coupling in inversion asymmetric magnets. Such a large antisymmetric exchange has been obtained at interfaces between heavy metals and transition metal ferromagnets, resulting in spin spirals and nanoskyrmion lattices. Here, using relativistic first$-$principles calculations, we demonstrate that the magnitude and sign of DMI can be entirely controlled by tuning the oxygen coverage of the magnetic film, therefore enabling the smart design of chiral magnetism in ultra$-$thin films. We anticipate that these results extend to other electronegative ions and suggest the possibility of electrical tuning of exotic magnetic phases.

\end{abstract} 

\maketitle

 \fontsize{10.5}{10}\selectfont Systems with broken spatial inversion symmetry present a fascinating playground for the design of thought-intriguing ferroic behaviors \cite{velev2011} and the emergence of unexpected transport phenomena \cite{hwang2012}. In magnetic solids lacking inversion symmetry, chiral magnetism can appear, which has been attracting increasing interest in the past ten years, with the direct observation of spin spirals and skyrmions in bulk crystals such as MnSi \cite{mulh2009} or Fe$_{0.5}$Co$_{0.5}$Si \cite{yu2010}. In these systems, the bulk inversion asymmetry results in the appearance of a Dzyaloshinskii$-$Moriya interaction \cite{dzyaloshinskii,moriya,fert80} (DMI) that gives rise to these chiral spin textures \cite{bogdanov2001,bogdanov20012,rossler2006}. \par

Another appealing class of systems in which chiral magnetic states have been recently unveiled are magnetic films and multilayers with interfacial inversion symmetry breaking. These systems are usually composed of an ultrathin (1 to 2 monolayers) layer of magnetic transition metal material (Fe, Mn etc.) deposited on top of a heavy metal (HM) substrate, such as Ir, Pd, W etc. Spin spirals were first observed by Bode \textit{et al.} \cite{bode2007} on Mn/W(110) and shortly later extended to Mn/W(100) \cite{ferriani2008}, Cr/W \cite{santosCond2008, zimmermanPRB-2014}, Fe/Ir \cite{heinze2011,mentzel2012} and Pd/Fe/Ir \cite{romming} systems. In contrast to the bulk crystals mentioned above where chiral states are obtained within a certain range of magnetic field and temperature, all these surfaces display spontaneous spin spirals and even a skyrmion crystal phase \cite{heinze2011} at low temperature. Numerical studies have demonstrated the seminal role of spin$-$orbit coupling in the form of DMI in the emergence and stabilization of these chiral magnetic textures.
\par
\vspace{0.2cm}
These magnet/heavy$-$metal interfaces are currently being intensively investigated from the standpoint of mainstream spintronics with the recent development of spin$-$orbit torques \cite{miron2011,liu-science-2012}. Indeed, the large current$-$driven torques arising from strong interfacial spin$-$orbit coupling \citep{manchon2008,manchon20081,manchon20082,freimuth2014,freimuth20141} in such ultrathin asymmetric structures constitute a promising means to electrically control of the magnetization direction and domain wall motion. In a recent work, Heide $et$ $al$. \cite{HeidePRB:2008} showed that interfacial DMI can alter the structure of magnetic domain$-$walls favoring N\'eel over Bloch configuration and leads to a stabilization sustaining a higher domain$-$wall speed \cite{thiaville2012}. Combined with spin$-$orbit torques, such distorted domain$-$walls are expected to display very large velocities, propagating against the electron flow \cite{emori2013,ryu2013}. These N\'eel walls have been recently identified using spin$-$polarized low$-$energy electron microscopy in asymmetric multilayers involving Ni/Ir and Ni/Pt interfaces \cite{chen-prl-2013,chen-natcom-2013} or using NV center magnetometry in AlO$_{x}$/Co/Pt trilayers \cite{Tetienne2014}. Alternative attempts have been engaged to validate and quantify the magnitude of DMI in these systems by analyzing the asymmetric creep motion of these domain walls \cite{hrabec2014,je2013,pizzini2014}. In the light of these recent achievements, the design of materials displaying large DMI constitutes a major goal in the development of current$-$controlled chiral domain walls, spin spirals and skyrmions \cite{Torrejon2014}.\par

In this work, we propose a new concept to fine$-$tune DMI by manipulating the degree of electronic asymmetry of a heavy metal/ferromagnet thin film through the adsoption of electronegative ions at the surface.
We validate this proposal by demonstrating that the magnitude and sign of the DMI of asymmetric ultrathin films can be widely tuned by controlling oxygen coverage. Although we demonstrate this effect in detail for oxygen, there is no doubt that it is much more general and should also apply to other ions whose electronegativity is larger than that of the heavy metal substrate, e.g. C, N, F, Cl, Br, I etc., or even to manipulations by an external electric field. Besides DMI, we explain in addition how other magnetic interactions, i.e. exchange interaction and the magnetocrystalline anisotropy energy (MAE), at such
transition$-$metal interfaces are modified in the presence of oxygen. Since complex magnetic structures such as, for example, skyrmions depend on a fine balance of these interactions, electronegative ions at interfaces may provide new opportunities to stabilize such structures. To demonstrate the effect, we consider a bilayer composed of a monolayer (ML) of Fe deposited on Ir(001) covered by O atoms of different coverage.
 In particular, we observe that DMI changes sign when the coverage exceeds 50\%. This study reveals that in realistic systems capped by an oxide, such as
MO$_{x}$/FM/HM (FM a ferromagnet and MO$_{x}$ might be MgO$_{x}$, AlO$_{x}$, CoO$_{x}$ etc.), DMI can be tuned by changing the oxidation conditions of the capping layer, offering a convenient way of control.  \\

{ \textbf{Methods}} \\

{ \textbf{First$-$principles calculations}} \\

\begin{figure} 
\includegraphics[width=0.7\textwidth ]{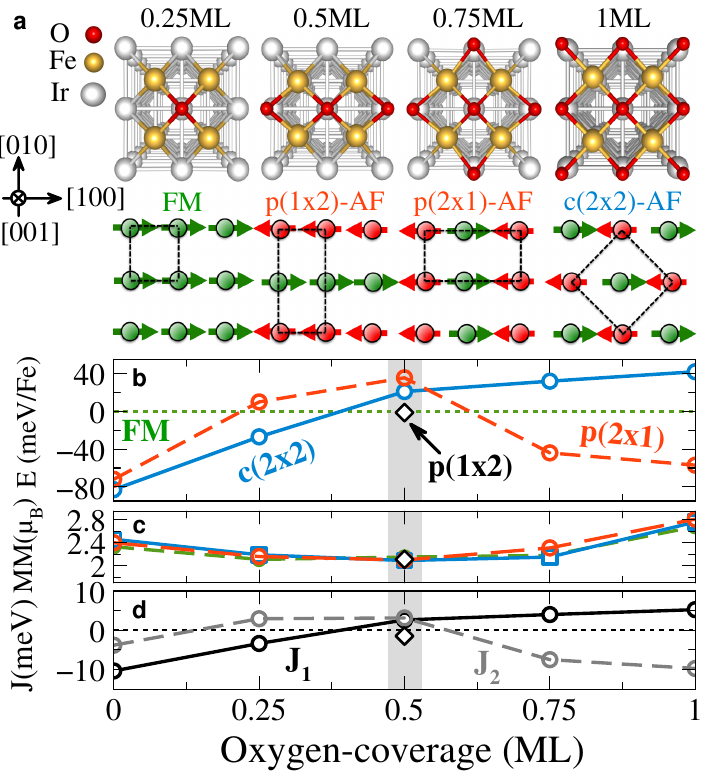}
\caption{\fontsize{10}{10}\selectfont \textbf{ Competing magnetic exchange interactions and geometrical
frustration in O/Fe/Ir(001)}. (a) Sketch of $p(2\times 2)$ unit cell centered with respect to the oxygen adlayer coverage (upper row) and the magnetic structure (lower row). (b) Total energy  per Fe
atom relative to the FM state of each oxygen
coverage, (c) magnetic moment of Fe atom, size of magnetic
moments of O and Ir are less then 0.1 $\mu_{B}$, and (d) exchange constants as a function of O coverage on Fe/Ir(001).
\label{fig1}}
\end{figure}

 We have determined the electronic, structural and magnetic properties of O/Fe/Ir(001) system by performing density$-$functional theory (DFT) calculations in the local density approximation (LDA) \cite{perd-zunger-LDA} to the exchange correlation functional, using the full$-$potential linearized augmented plane wave (FLAPW) method in film geometry \cite{Wimmer1981} as implemented in the F{\footnotesize LEUR} code \cite{Fleur}.  The system was modeled by an asymmetric film consisting of a monolayer of Fe in fcc stacking on seven layers of Ir representing the Ir(001) substrate. The Fe surface was covered with O at five different coverages $\Theta$=0, 0.25, 0.5, 0.75 and 1 ML (see \textcolor{blue} {Fig. \ref{fig1}(a)}), defined relative to the number of Fe atoms per unit area. All atomic surface structures obey the $C_\textrm{4v}$ point symmetry with the exception of $\Theta=0.5$, here the symmetry is reduced to $C_\textrm{2v}$ in the energetically preferred adsorption geometry (see \textcolor{blue} {Fig. \ref{fig1}(a)}). Thus $\Theta$=0, 1ML represent the plain Fe/Ir bilayer and the limit of full coverage, respectively. All calculations are carried out in the $p$(2$\times$2) unit cell, using the experimental Ir bulk lattice constant of $a_{\textrm{bulk}}$=3.84 \AA. For each of the five oxygen coverages the atom positions of the O, Fe and three topmost Ir layers where optimized for four different collinear states: the FM (ferromagnetic) state and the row$-$wise $p(2\times1)-$, $p(1\times2)-$, and checkerboard $c(2\times2)-$AFM (antiferromagnetic) states, as displayed in \textcolor{blue} {Fig. \ref{fig1}(a)}. Then, the global magnetic ground state was explored starting for each coverage with the atomic structure of the lowest$-$energy collinear magnetic configuration by calculating self$-$consistently the total energy for homogeneous spin$-$spirals. The energy of these spin spirals of wave vectors $\textbf{q}$, taken along high$-$symmetry directions of the two$-$dimensional Brillouin zone (2D$-$BZ), were calculated using the generalized Bloch theorem \cite{KurzPRB-2004}. On$-$top we added the spin$-$orbit coupling (SOC) by first$-$order perturbation theory \cite{Heide-phys-B} to evaluate the contribution of DMI to the total energy in the vicinity of the spin$-$spiral ground state obtained without SOC. The MAE is obtained for all O$-$coverages and all collinear states at relaxed geometry as total$-$energy difference for two different magnetization directions employing the force theorem. Note that high computational accuracy is required since energy differences between different magnetic orderings are tiny ($\sim$meV) in the present case.  We considered 512 and 1024 \textbf{k}$_{\parallel}-$points in the 2D$-$BZ for scalar relativistic, and the calculation with SOC, respectively. 

\vspace{0.75cm}
{ \textbf{Results} \par
{ \textbf{Magnetic ground states and spin interactions in O/Fe/Ir(001)} 
\vspace{0.3cm}

 In the case of low O coverage, namely for $\Theta$=0 and 0.25 ML, the Fe ML exhibits a clear antiferromagnetic behavior and adopts the checkerboard $c(2\times2)-$AFM order (see \textcolor{blue} {Fig. \ref{fig1}(b)}). For high O coverage, i.e. $\Theta$=0.75 and 1 ML, the Fe ML is again antiferromagnetic and displays the $p(2\times1)-$AFM configuration. Interestingly, for an ordered half$-$ML of oxygen, the $p(1\times2)-$AFM state is energetically favored over the $c(2\times2)-$AFM configuration by $\sim$22 meV/Fe ($p(2\times1)-$AFM by $\sim$36 meV/Fe), which itself lies 1.34 meV/Fe below the FM state (see \textcolor{blue} {Fig. \ref{fig1}(b)}). Within the accuracy of our calculations, the FM and the $p(1\times2)-$AFM configurations become almost degenerate. The extremely small energy difference between $p(2\times1)-$ and FM configurations at 0.5ML O coverage indicates a strong influence of the oxygen on the magnetic ordering and on the weakening of the AFM interaction due to hybridization with the Fe atoms. Indeed, Fe$-$O hybridization increases the Fe$-$Ir interlayer distance upon oxygen adsorption, and consequently changes the interaction between the Fe atoms and their hybridization with the Ir substrate. Most importantly, this particular coverage breaks the $C_\textrm{4v}$ symmetry with respect to the Ir coordination as compared to the oxygen$-$free case. It should be noted that atomic relaxations, which can strongly influence the magnetic ground state, are very important for the stabilization of the magnetic ordering. These results are in excellent agreement with the theoretical study at the LDA$-$PAW level by M\'aca et $al$. \cite{maca2013}, who predicted the same energetic orderings at 0.5ML as that in \textcolor{blue} {Fig. \ref{fig1}(b)}.

For coverages above 0.5ML, the magnetic ordering is strikingly different from the lower coverages of oxygen in which the $p(2 \times 1)-$AFM state becomes energetically favorable among all states. The FM configuration assumes an intermediate position between $p(2\times1)-$ and checkerboard $c(2\times2)-$AFM. The occurrence of this magnetic configuration has also been confirmed for 1ML \cite{maca2013}. This gives us confidence that we capture the correct trend in magnetic orderings already within the employed model of (2$\times$2) geometry. From the above results we can conclude that the AFM state is preferred over the FM exchange interaction in the whole coverage range of O except at 0.5ML in which the AFM and FM states are degenerate.

By mapping the results onto a two$-$dimensional (2D) classical Heisenberg model, we can determine the exchange constants between moments of Fe adatoms on Ir(001) substrate. The corresponding Heisenberg Hamiltonian that describes the magnetic interaction is given by $H_{\mathrm{ex}} = - \sum_{ij}  J_{ij} \vec{S_{i}} \cdot \vec{S_{j}}$ where $J_{ij}$ is the exchange interaction constant between two spins $\vec{S}$ at sites $i$ and $j$, and its sign determines whether FM ($J_{ij}>0$) or AFM ($J_{ij}<0$) spin alignment between $S_{i}$  and $S_{j}$ is energetically favored. Such a phenomenological model with fixed spin values depends weakly on the magnetic moment of Fe atoms with respect the O coverage and magnetic order \textcolor{blue} {Fig. \ref{fig1}(c)}. For a two$-$dimensional square lattice, the nearest neighbor (NN) and next$-$nearest neighbor (NNN) exchange constants $J_1$ and $J_2$, respectively, can be estimated from the total energies of the different magnetic configurations by mapping them to a Heisenberg model. The corresponding exchange constants are defined as \cite{maca2013}  \par
$J_{1}= \frac{1}{8}[\textit {E}_{c(2\times2)}^{\rm  tot}(\rm AFM)-\textit {E} ^{tot}(\rm FM)]$, \par
$J_{2}= \frac{1}{8}[\textit {E}_{p(2\times1)}^{\rm tot}(\rm AFM)-\textit {E}^{tot}(\rm FM)-4  \textit{J}_{1} ]$.

The coverage dependence of the exchange constant is shown in \textcolor{blue} {Fig. \ref{fig1}(d)}. The most remarkable feature is the dramatic change of the leading first$-$NN interaction $J_1$, from the strongly AFM coupling for the lower oxygen coverages to the FM for the higher coverages, roughly beyond 3/8 ML. Note that the variation of the first$-$ and second$-$NN exchange interactions mimic the energies of the magnetic ordering [see  \textcolor{blue} {Fig. \ref{fig1}(b)}].  Regarding the NNN exchange interaction $J_2$, we find that the sign has an oscillating behavior in the considered coverage range due to the competition of FM and AFM order beyond the next NN. The NNN interaction has the AFM character above 0.5ML. However, for lower coverages around $\Theta$=0.25 both $J_1$ and $\ J_2$ are of comparable magnitudes and become nearly degenerate at $\Theta$=0.5, which leads to pronounced frustration of the competing exchange interactions of these AFM states. \textit{Thus, we expect the possibility of a noncollinear ground state close to this transition point}. 

\vspace{0.5cm}
{ \textbf{Spin$-$spirals and Dzyaloshinskii$-$Moriya interaction} 
\vspace{0.4cm}

The small exchange interaction obtained at 0.5ML indicates that even small SOC can be a central ingredient in the stabilization of complex magnetic textures. This suggests extending the calculations by including spin spirals and relativistic effects (i.e. magnetic anisotropy and DMI), which can contribute significantly to the formation of chiral domain walls and skyrmions. In order to understand whether the formation of spiral spin texture is energetically favorable in O/Fe/Ir(001), we consider the energy of planar spin$-$spirals with wavevector {\bf q} along the high symmetry lines in the irreducible 2D$-$BZ.
Considering the propagation vector {\bf q} along the $\overbar{\Gamma X}$ direction, the magnetic moment  of an atom at site \textbf{R} points in the direction $\hat{\mathrm{\textbf{s}}}(\textbf{R})=[\mathrm{cos}(\textbf{q} \cdot \textbf{R}), 0 ,\mathrm{sin}(\textbf{q} \cdot \textbf{R})]$. The magnitude of the wave vector {\bf q} is then varied from {\bf q}=0 (FM state) to $q$= $\pm$0.5$\frac{2\pi}{a'}$ $[(\textbf{q} \cdot \textbf{R})=90^{\circ}]$, where $a'$ is the lattice constant of the $p(2\times2)$ unit cell $a'=a_\textrm{Ir-bulk}\sqrt{2}$, which is the maximum of {\bf q}$-$vector along $\overbar{\Gamma X}$ direction in the corresponding 2D$-$BZ of $p(2\times2)$ unit cell (the corresponding 2D$-$BZ is smaller by a factor 2 compared to the $p(1\times1)$ unit cell). Notice that by the calculation of the spin spiral we can cover all possible magnetic phases that can be described by a single {\bf q}$-$vector. For instance, the high$-$symmetry points $\overbar{\Gamma}$, $\overbar{\rm M}$, and $\overbar{X}$, which correspond to the previously discussed collinear states: FM,  checkerboard $c(2\times2)-$AFM, and row$-$wise $p(2\times 1)-$ or  $p(1\times2)-$AFM state, respectively, are attainable  by extending the spin$-$spiral vector until {\bf q}$ =\pm \frac{2\pi}{a'}\sqrt{2}$ or {\bf q}$ = \pm \frac{2\pi}{a'}$ in the 2D$-$BZ. \par

\begin{figure}[t!]
\includegraphics[width=0.65\textwidth ]{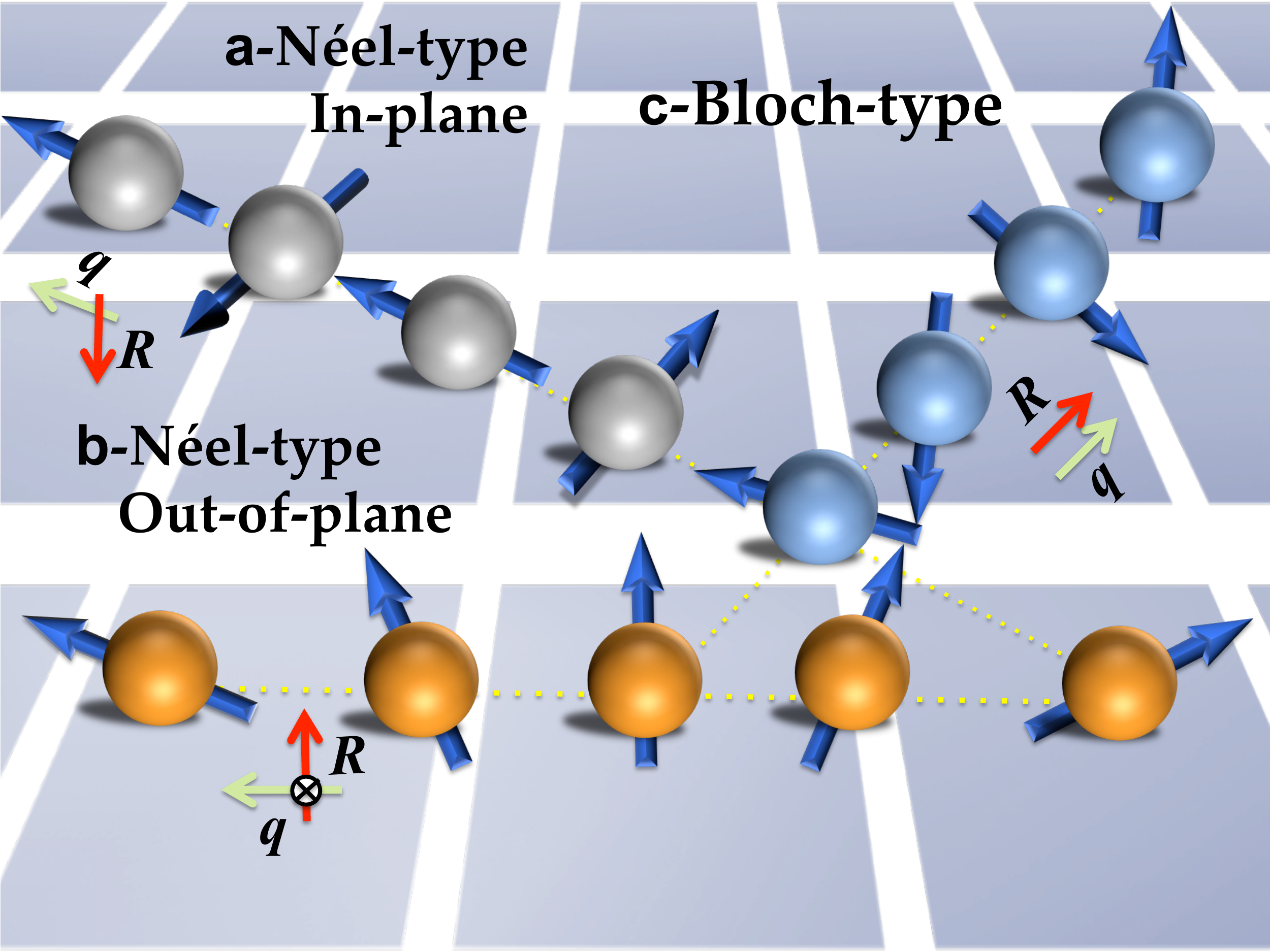}
\caption{\fontsize{10}{10}\selectfont \textbf{Schematic representation of spin$-$spirals with different propagation directions $\textbf{q}$ and spin$-$rotation axis $\bf{R}$}. (a,b) N\'eel$-$type in$-$plane and out$-$of$-$plane $(\bf{R \perp q})$, respectively and (c) Bloch$-$type with $(\bf{R \parallel q)}$. Note that the DMI vanishes for N\'eel$-$in$-$plane and Bloch$-$type due to symmetry arguments.
\label{fig2}}
\end{figure}
\vspace{0.2cm}

 When considering the SOC for a spin$-$spiral state, two additonal energy contributions appear. The first contribution is the symmetric MAE, of the form $E=\sum_{i} K_{i} \mathrm{sin}^2 \varphi_{i}$ with the magnetic anisotropy energy $K_{i}$ and the angle $\varphi_{i}$ between the easy axis at site $i$ and the magnetization axis. The second contribution is the antisymmetric DMI, which in terms of a spin model is of the form $E_{\mathrm{DM}}=\sum_{i,j} \mathrm{\bold{D}_{ij}} \cdot (\bold{S}_{i} \times \bold{S}_{j})$, where $\mathrm{\bold{D}_{ij}}$ is the DM vector, which determines the strength and sign of DMI. Note that the spin$-$wave configurations depend on the different combinations of the rotation axis $\bf{R}$ with the wave$-$vector propagation $\bf{q}$, which are essentially important in defining the sign and direction of DMI. Based on the symmetry analysis predicted by Moriya \cite{moriya}, if we consider the N\'eel$-$type out$-$of$-$plane configuration along the high$-$symmetry direction ([100] or [010]), then the $\bf{q}$ vector should be perpendicular to $\bf{R}$ $(\bf{R \perp q})$ with no mirror plane due symmetry breaking at the interface. In this case we conclude that the $\mathrm{\bold{D}_{ij}}$ vector is oriented in plane and normal to the $\bf{q}$ vector (see \textcolor{blue}{Fig. \ref{fig2}(b)}). However, because of the preserved
mirror$-$plane according to the symmetry argument \cite{moriya} the DMI term must vanish for both configurations, Bloch$-$type spin$-$spiral with $(\bf{R \parallel q)}$ and N\'eel$-$type in$-$plane with $(\bf{R \perp q})$ (see \textcolor{blue}{Fig. \ref{fig2}(a, c)}). 

\begin{figure}[t!]
\includegraphics[width=0.7\textwidth ]{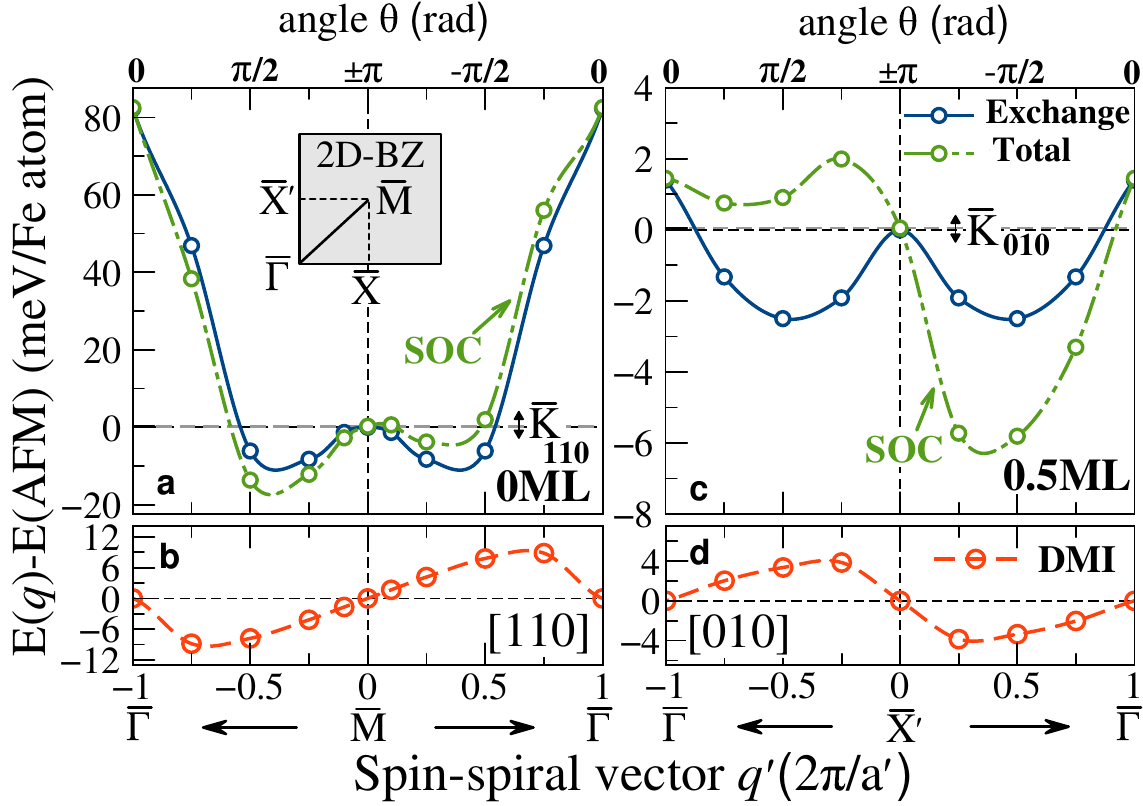}
\caption{ \fontsize{10}{10}\selectfont \textbf{Energy dispersion of N\'eel spin$-$spirals in Fe/Ir(001) and O/Fe/Ir(001)}. (a) Spin$-$spiral dispersion energy including SOC for Fe/Ir(001)($\Theta$=0ML) propagating along $\overbar{\Gamma \rm{M}}$ direction of the 2D$-$BZ (see inset) (c) for O/Fe/Ir(001) ($\Theta$=0.5ML) along $\overbar{\Gamma X'}$ direction. In (a) and (c) the energies are given relative to the AFM state. 
Contributions from Heisenberg exchange (blue line), (b) and (d) Dzyaloshinskii$-$Moriya interaction (SOC) (red line), and their sum (green line). According to the micromagnetic model, the energy dipersion of the spin spirals is vertically shifted with respect to the antiferromagnetic state due to the magnetocrystalline anisotropy $\overline{K}$. 
\label{fig3}}
\end{figure}

\vspace{0.2cm}
Starting from the FM configuration, we calculated the energy dispersion of N\'eel spirals $E({q'})$ including the SOC for ${q'}$ along the high$-$symmetry directions $\overbar{\Gamma \rm{M}}$ and  $\overbar{\Gamma X'}$ [\textcolor{blue}{Fig. \ref{fig3}(a, c)}] for Fe/Ir(001) and O/Fe/Ir(001)$-$($\Theta$=0.5), respectively. Because of the $C_\textrm{4v}$ (fourfold) symmetry$-$breaking at 0.5ML, the $\overbar{\Gamma X}$ and $\overbar{\Gamma X'}$ directions in the 2D$-$BZ are no longer equivalent [see the inset of \textcolor{blue}{Fig. \ref{fig3}(a)}]. When the energy $E({q'})$ along the high symmetry lines of 2D$-$BZ is lower than any of the collinear magnetic phases studied previously, the system most likely adopts an incommensurate spin$-$spiral magnetic ground$-$state structure.\par

For Fe/Ir(001), when excluding the SOC from the calculations, the energy dispersion is symmetric around the $\Gamma$ and M points in the $\overbar{\Gamma \rm{M}}$ direction with a local energy minimum at ${q'}$=$\pm 0.42\frac{2\pi}{a'}\sqrt{2}$, located about 11 meV/Fe atom below the AFM state (${q'}$=0) (see blue line in \textcolor{blue}{Fig. \ref{fig3}(a)}). The energies are given relative to the $c(2\times2)-$AFM state. By considering a deviation $\delta {\bf q}$ from ${\bf q}_{\rm{AFM}}$, the wave vector ${q'}$ can be written in the form [$\frac{2\pi}{a'}+\delta {\bf q}$] and [$\frac{2\pi}{a'}\sqrt{2}+\delta {\bf q}$], which corresponds to AFM states, $\overbar{X}$ and $\overbar{\rm{M}}$ points, respectively. Thus, the period length of the AFM (${q'}$=0) is defined to be infinite in notion of the magnetic the magnetic supercell ($\lambda^{-1}=\pm \infty$). The positive and negative values of vector ${q'}$ refer to a counter$-$clockwise and clockwise rotating spiral, respectively. However, for larger angles $\rm{\theta}$ between adjacent moments, i.e., for a shorter wave length $|\lambda|$ of the spin$-$spiral, the energy rises faster due to the strong AFM$-$NNN exchange coupling and spin$-$spiral configurations become unfavorable. Note that this picture changes significantly in the presence of SOC, since the DMI is strong enough to compete against the Heisenberg exchange and MAE averaged over the pitch of the spiral ($\overline{K}$) that favors an in$-$plane easy axis (\textcolor{blue}{Fig. \ref{fig3}(a)}, see also \textcolor{blue}{ Table \ref{tab1}}). Interestingly, the SOC breaks the inversion symmetry between left and right$-$hand rotating strutcture (see \textcolor{blue}{Fig. \ref{fig3}(a)}), and leads to an antisymmetric exchange DMI that contributes for small $q'$ by a linear term in $\lambda^{-1}$ to the total energy (\textcolor{blue}{Fig. \ref{fig3}(b)}), giving rise to a unique direction of the vector spin chirality of our magnetic structure \cite{mentzel2012}. As a result, the strong DMI lifts the degeneracy of the spin$-$spiral ground state in favor of the left$-$handed (cycloidal) spin$-$spiral with a significant energy gain of about 17 meV per Fe atom, which leads finally to a spin$-$spiral ground state with a pitch of $|\lambda_0|$$\approx$0.5 nm. Apart from the different magnetic ground state this behavior is similar to the one found for Fe/Ir(111), where the large DMI interaction can lead to the formation of skyrmions \cite{heinze2011,dupe}. \\

{ \textbf{Oxygen$-$enabled control of Dzyaloshinskii$-$Moriya interaction}} 
\vspace{0.5cm}

However, the situation changes dramatically in the case of 0.5ML oxygen coverage, as displayed in \textcolor{blue}{Fig. \ref{fig3}(c)}. For both directions, the energy difference between N\'eel spirals and $p$(1$\times$2)$-$AFM state is significantly decreased indicating a weakening of the NN exchange couplings $J_{1}$ and $J_{2}$. Indeed, the FM and $p(1\times2)-$AFM states are almost degenerate for this particular coverage due to the breaking of the $C_{4}$ symmetry as discussed above. For smaller $|\lambda|$ close to the FM state, the spin$-$spirals are energetically unfavorable with respect to the AFM state. Contrary to the clean Fe/Ir(001), the DMI is sufficiently strong to stabilize a chiral magnetic ground state with the right$-$rotational sense. As we can see in \textcolor{blue}{Fig. \ref{fig3}(c, d)}, an energy minimum of $\approx$ 6.3 meV/Fe atom now appears for a right$-$handed chirality with a pitch of $|\lambda_{0}|$=0.75 nm. This energy minimum $E_{0}(q')$ at ${q'}$=$\pm 0.36\frac{2\pi}{a'}$ is $\approx$ 7.8 meV/Fe atom lower than the FM configuration. For  the same coverage (0.5ML) but along $\overbar{\Gamma X}$ direction, the DMI is also strong enough to compete against the other interactions to stabilize the spin$-$spirals (not shown). Note that the sign of DMI for Fe and Ir atoms at the Fe$-$Ir interface is the same, irrespective whether the frozen spiral propagates along [010] or [100] direction.\par

\begin{table}[b!]
\begin{ruledtabular}
\caption{\label{tab1}
\fontsize{10}{10}\selectfont  The magnetocrystalline anisotropy energy (MAE) in meV per Fe atom for different oxygen coverages deposited on Fe/Ir(001) surface. Negative (positive) values of the MAE indicate an out$-$of$-$plane (in$-$plane) magnetization.
}
\begin{tabular}{c c c c c c }
& Coverage (ML) &  FM&$c$(2$\times$2)$-$AFM& $p$(2$\times$1)$-$AFM & $p$(1$\times$2)$-$AFM  \\
 \hline

& 0.00&1.43 & $-$0.56 &	{ } 2.15 & { } 2.15\\
&0.25&	0.69&	$-$1.21&{ }	1.44&{ } 1.44\\
&0.50&1.65&	$-$1.77 &	$-$0.56 & $-$0.10 \\  
&0.75 &	1.17 &	$-$0.03&{ }	1.22&$-$0.05 	\\
&1.00& 3.65 &	{ } 1.37 	 &{ } 1.43 &$-$0.50\\

\end{tabular}
\end{ruledtabular}
\end{table} 

Another striking feature is the change of sign of DMI with respect to the clean reference Fe/Ir(001), as can be seen in \textcolor{blue}{Fig. \ref{fig3}(d)}, demonstrating the sensitivity of DMI and the exchange interaction to the oxygen overlayer. Layer$-$resolved calculations (not shown) indicate that the sign and magnitude of DMI are mainly ascribable to the large SOC of Ir surface and do not depend significantly on the 3$d-$Fe layer. However, although O$-$2$p$ states are weakly spin$-$orbit coupled, they can dramatically modify the band filling and 3$d-$Fe 5$d-$Ir hybridization through symmetry breaking and charge transfer \cite{kashid,BornemannPRB-2012}. A similar effect has been reported for O/Gd(0001) interface by Krupin $et$ $al.$ \cite{kurpinPRB2005}. This effect is illustrated by the increase of the Fe$-$Ir interlayer distance upon O adsorption, which weakens Ir$-$Fe hybridization at the interface and hence, reduces the exchange interaction, as seen in \textcolor{blue}{Fig. \ref{fig3}(a), (c)}. \par

\begin{figure}[b!]
\includegraphics[width=0.65\textwidth ]{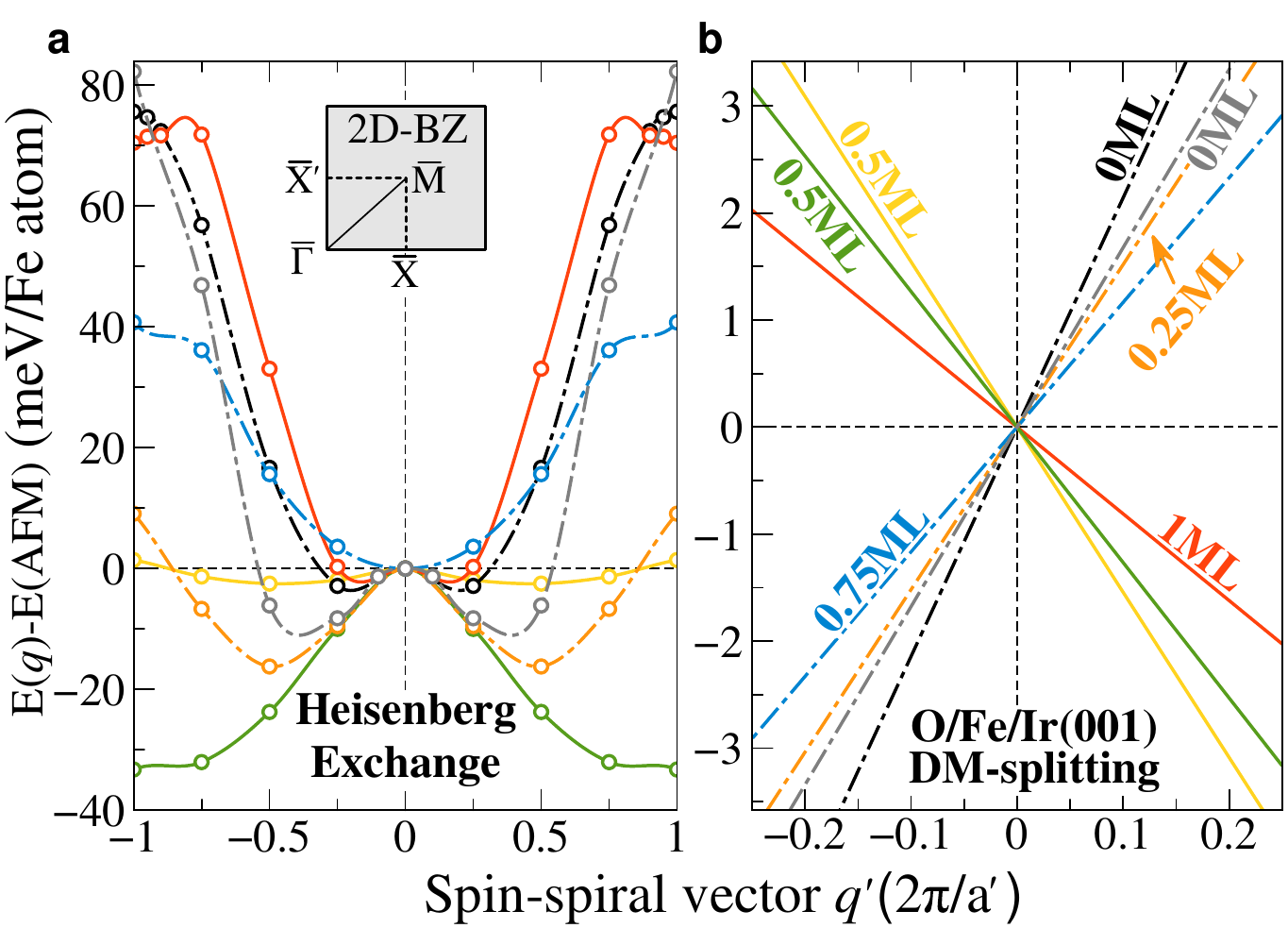}
\caption{ \fontsize{10}{10}\selectfont \textbf{Impact of the surface oxidation on the magnetic interactions in Fe/Ir(001)}. (a) The isotropic exchange$-$interaction energy $E_{0}(q')$, (b) the Dzyaloshinskii$-$Moriya interaction energy $\Delta E_{\rm DM}(q')$ for O/Fe/Ir(001) as a function of oxygen coverage  $\Theta$ varying between 0 and 1ML. The results shown in (a) are obtained
without SOC. (b) The slope of this odd part at $E(q')$=0 determines the sign and magnitude of DMI.
The green and yellow lines for 0.5 ML correspond to ($\overbar{\Gamma}\leftarrow\overbar{\rm X}\rightarrow \overbar{\Gamma}$)  and ($\overbar{\Gamma}\leftarrow\overbar{\rm X}'\rightarrow\overbar{\Gamma}$) directions, while black and gray lines for 0ML correspond to ($\overbar{\Gamma}\leftarrow \overbar{\rm X}\rightarrow \overbar{\Gamma}$) and ($\overbar{\Gamma} \leftarrow \overbar{\rm M} \rightarrow \overbar{\Gamma}$), respectively.
\label{fig4}}
\end{figure}

To get a better understanding in how the magnetic interactions are influenced by the presence of oxygen, we provide a complete picture of oxygen impact on the Fe exchange coupling and DMI interaction as a function of O coverage in \textcolor{blue}{Fig. \ref{fig4} (a), (b)}. First, the exchange constants (mainly $J_1$) are greatly reduced from 0ML to 0.5ML and the energy differences to other metastable configurations are significantly reduced. For instance, for $q'$=0.75, the energy of the spin$-$spiral is reduced from 58 meV/Fe atom for 0ML to $-$31 meV/Fe atom for half oxygen coverage.
This indicates that the presence of oxygen even in the lower coverage can alter significantly the nature of the magnetic interaction of the entire system. However, for high coverage above 0.5ML, the energy
difference to the Ne\'el state rises again with decreasing the period$-$length  $\lambda$ to about 72 meV/Fe atom for full monolayer, indicating a strong AFM$-$NNN
exchange interaction. We should note that the spin$-$spiral state remains the global energy minimum for all coverages in the presence of SOC, the only exception being 1ML where AFM is the ground state.

More intriguing is the behavior of DMI with respect to the O coverage. As displayed in \textcolor{blue}{Fig. \ref{fig4}(b)}, the sign of DMI which determines the handedness in Fe/Ir(001) interface, shows an oscillating behavior in the sequence of O coverage 0.25($+$), 0.5($-$), 0.75($+$), and 1ML($-$), with a remarkable DMI reduction for half and full coverage. Since DMI is obtained within the first$-$order perturbation theory in SOC, it can be readily connected to the modification of the Fermi energy induced by the SOC. This energy shift is defined as $\delta \varepsilon = (\varepsilon_{\mathrm F} -\varepsilon_{\mathrm F}^0)$, where
$\varepsilon_{\mathrm F}$ and $\varepsilon_{\mathrm F}^0$ are the Fermi energies calculated with and without SOC, respectively. This energy shift $\delta \varepsilon$ is reported in \textcolor{blue}{Fig. \ref{fig5}(c)} for a spin$-$spiral of wave vector $q'$=0.25. We observe that $\delta \varepsilon$ exhibits an oscillatory behavior when varying the oxygen coverage, consistent with the sign of DMI (\textcolor{blue}{Fig. \ref{fig5}(b)}). This explains the sensitivity of the magnitude and sign of $E_{\mathrm{DMI}}$ on the details of hybridization between the transition$-$metal 3$d-$Fe,  5$d-$Ir, and 2$p-$O orbitals upon adsorption of the oxygen atoms, as mentioned above. This observation calls for a deeper analysis of the influence of O adsorption on the electronic structure of the system near the Fermi energy. \par

In order to further understand the trends found in \textcolor{blue} {Fig. \ref{fig5}(b)} and \textcolor{blue}{5(c)},  it is instructive to analyze the charge transfer and induced dipole moment (Dm), when O is adsorbed on Fe/Ir(001) interface, as shown in \textcolor{blue} {Fig. \ref{fig5}(a)} and \textcolor{blue}{5(b)}.
 To understand the nature of the bonding, the charge density differences are calculated using $\Delta \rho(\textbf{r})=\rho_{\mathrm{tot}}(\textbf{r})-[\rho_{\mathrm{Ir}}(\textbf{r})+\rho_{\mathrm{Fe}}(\textbf{r})+\rho_{\mathrm{O}}(\textbf{r})]$, where $\rho_{\mathrm{Ir}}(\textbf{r})$, $\rho_{\mathrm{Fe}}(\textbf{r})$, $\rho_{\mathrm{O}}(\textbf{r})$, and $\rho_{\mathrm{tot}}(\textbf{r})$ are the charge density distributions of the Ir surface, Fe layer, O adatoms, and the conjugate system, respectively, each in the precise position they adopt in the adsorption system. The dipole moment $\mu(z)$ induced by O adsorption can be obtained by integrating  the half$-$cell volume along the $z-$direction $\mu(z)=\frac{1}{n}\int_{-c}^{-c+z/2} {z' \Delta \overline{\rho}(z')} dz'$, where $z$/2 is half the length of the supercell, $c$ is the distance from the tompost layer to the middle of the slab, and $n$ is the number of adsorbed oxygen atoms per unit cell. Note that the planar averaged charge density difference $\Delta\overline{\rho}(z')$ is $\Delta \rho(\textbf{r})$ integrated over lateral coordinates $x$ and $y$ for each $z$ plane, as shown in \textcolor{blue} {Fig. \ref{fig5}(a)}.

\begin{figure}[h!]
\includegraphics[width=0.75\textwidth ]{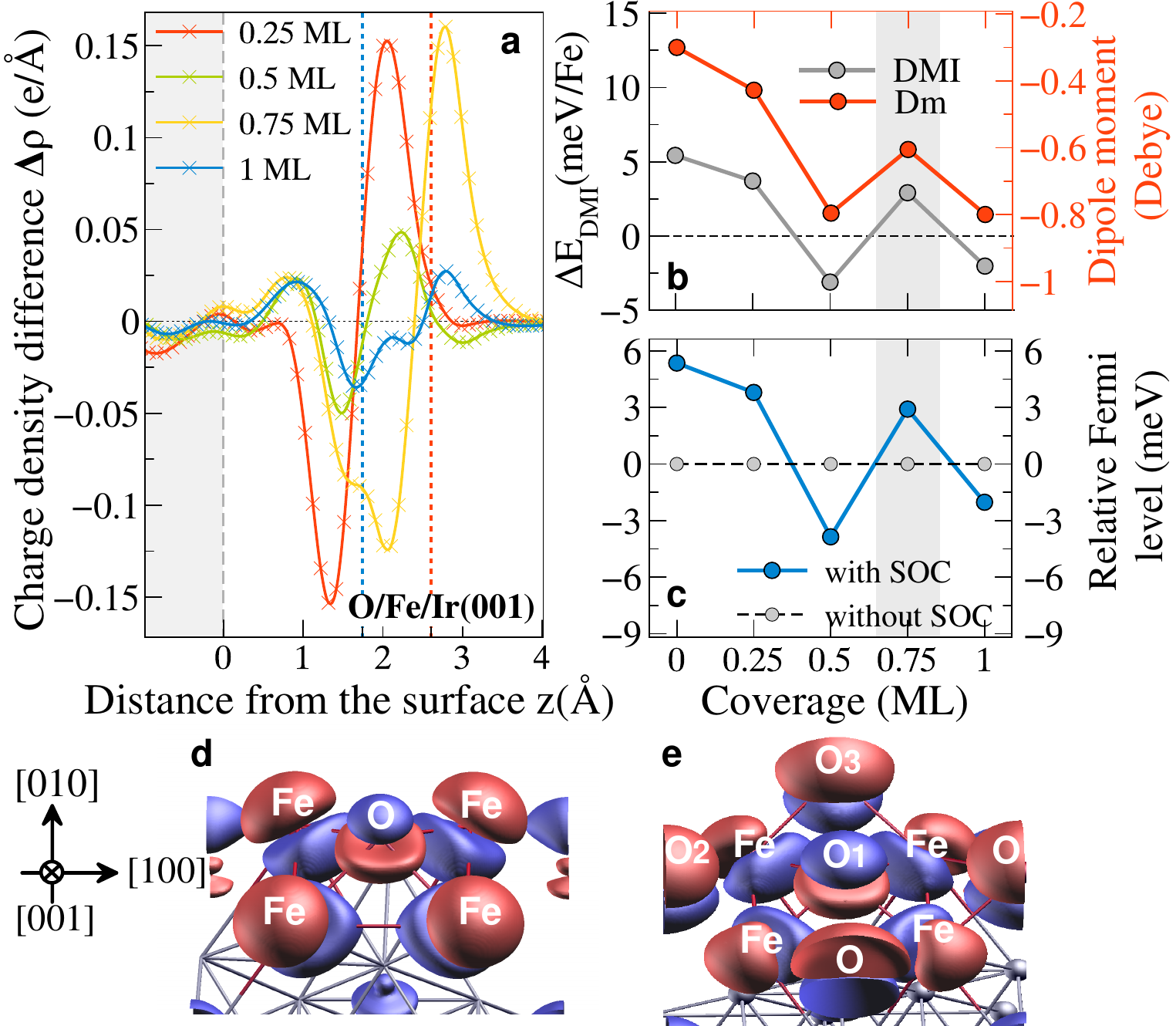}
\caption{\fontsize{10}{10}\selectfont  \textbf{Correlation between Dzyaloshinskii$-$Moriya interaction (DMI) and the electric surface dipole moment (Dm)}. (a) Planar averaged charge density difference $\Delta \rho(z)$ for O adsorption on Fe/Ir(001) at different coverages. The dashed red, blue, and gray lines show the approximate equilibrium positions of the O, Fe and tompost Ir atoms, respectively, on the fully relaxed surface.  O coverage dependence of the change in the surface dipole moment (Dm), DMI (b) and relative Fermi energy (c). Isosurface plot of the charge density difference $\Delta \rho(\textbf{r})$ for 0.25 (d) and 0.75ML (e),  blue and red isosurface plots ($ 1.2\times 10^{-2}e/${\AA}$^3$) represent the region of charge accumulation and depletion, respectively. }
\label{fig5}
\end{figure}

 We find interesting trends across the oxygen coverage considered in our calculations for which the strength and sign of DMI clearly correlate with the charge transfer and the induced electrostatic dipole moment.
The electron transfer from the Fe layer to the oxygen adatom is significant and reflects the high electronegativity of oxygen. Indeed, the charge transfer depending on the oxygen coverage, it affects the position of Fe$-3d$ levels with respect to the Fermi energy and controls by hybridization the band filling of the Ir$-5d$ orbitals.  Note that the lower surface dipole moment and consequently lower DMI at 0.75ML is mainly due
to the strong reduction of the repulsive electrostatic interactions between O adatoms. As a consequence, there is partial electron transfer back to Fe layer which leads to surface depolarization. This fact is clearly visible at the neighbors O (2$-$3) atoms in \textcolor{blue} {Fig. \ref{fig5}(e)} compared to 0.25 ML in \textcolor{blue}{Fig. \ref{fig5}(d)}. A similar effect has been reported for oxygen on transition metals by Zhang $et$ $al.$ \cite{stampflPRB}. Therefore, the occurrence of a charge accumulation and depletion on oxygen adatoms is very important in the context of the work function (see \textcolor{blue}{Fig. \ref{fig5}(c)}). Indeed, it plays an essential role to elucidate the tendency of the induced electric dipole moment and, thereby, the sign and magnitude of DMI change. \newpage

{ \textbf{Discussion}} 

In summary, we have explored the impact of the surface oxidation on the magnetic texture in transition metal interfaces. In particular, we show that the DMI in Fe/Ir(001) interface can be controlled by oxygen adsorption, which changes the wavelength of the spin spiral and its rotational sense. Note that for an oxygen coverage of 0.5ML, DMI is sufficiently strong to stabilize a chiral magnetic ground state
of the right$-$rotating cycloidal spin$-$spiral. In addition, we demonstrated that the sign and strength of DMI depend strongly on the electric surface
dipole moment induced by charge transfer and related hybridization between 2$p-$O and (3$d-$Fe,  5$d-$Ir)  states around the Fermi level. We anticipate that these effects are not limited to oxygen but can be extended to other electronegative ions such as C, N, F, etc. Furthermore, this work suggests that the modification of the interfacial electrical polarization through controlled oxidation of capping insulating oxides (MgO$_{x}$, AlO$_{x}$ etc.) or, to some extent, gate voltages could efficiently tune the magnetic state of the underlying magnet. This demonstration paves the way towards the design of chiral magnetic properties through interface engineering, which offers promising perspectives in terms of electrical control of the magnetic chirality. \par

 \vspace{1cm}

{ \textbf{Acknowledgments}} \\

A.B. and A.M. acknowledge financial support from the King Abdullah University of Science and Technology (KAUST).  We acknowledge computing time on the supercomputers SHAHEEN, NOOR, and SMC at KAUST Supercomputing Centre and JUROPA at the J\"ulich Supercomputing Centre (JSC). \\

{\textbf{Author contributions}} \\

A.B. and G.B. carried out the density functional theory calculations. All authors discussed the theoretical data and contributed to writing and preparing the paper. \\

{ \textbf{Competing financial interests}} \\

The authors declare no competing financial interests.


\begin{thebibliography}{1}
\fontsize{10}{10}\selectfont{
\bibitem{velev2011} Velev, J. P., Jaswal, S. S. \& Tsymbal, E. Y. Multi-ferroic and magnetoelectric materials and interfaces. Phil. Trans. R. Soc. A {\bf369}, 3069 (2011).
\bibitem{hwang2012} Hwang, H. Y., Iwasa, Y., Kawasaki, M., Keimer, B., Nagaosa, N., Tokura \& Y. Emergent phenomena at oxide interfaces. Nat. Mater. {\bf 11}, 103 (2012).

\bibitem{mulh2009} M\"uhlbauer, S., Binz, B., Jonietz, F., Pfleiderer, C., Rosch, A., Neubauer,  A., Georgii, R. \& B\"oni, P. Skyrmion Lattice in a Chiral Magnet. Science {\bf 323}, 915 (2009).

\bibitem{yu2010} Yu, X. Z., Onose, Y., Kanazawa, N., Park, J. H., Han,  J. H., Matsui, Y., Nagaosa N. \&  Tokura, Y. Real-space observation of a two-dimensional skyrmion crystal. Nature {\bf 465}, 901 (2010).

\bibitem{dzyaloshinskii} Dzyaloshinskii, I. E. A thermodynamic theory of weak ferromagnetism of antiferromagnetics. Sov. Phys. JETP {\bf5}, 1259 (1957).

\bibitem{moriya} Moriya, T. Anisotropic Superexchange Interaction and Weak Ferromagnetism. Phys. Rev. {\bf120}, 91 (1960).

\bibitem{fert80} Fert, A. \& Levy, M. Role of Anisotropic Exchange Interactions in Determining the Properties of Spin-Glasses. Phys. Rev. Lett. {\bf 44}, 1538 (1980). 

\bibitem{bogdanov2001} Bogdanov, A. N. \& R\"o\ss{}ler, U. K. Chiral Symmetry Breaking in Magnetic Thin Films and Multilayers. Phys. Rev. Lett. {\bf87}, 037203 (2001).

\bibitem{bogdanov20012}  Bogdanov, A. \&  Hubert, A. Thermodynamically stable magnetic vortex states in magnetic crystals.  J. Magn. Magn. Mater. {\bf138}, 255 (1994).

\bibitem{rossler2006} R\"o\ss{}ler, U. K., Bogdanov, A. N. \& Pfleiderer, C. Spontaneous skyrmion ground states in magnetic metals. Nature {\bf 422}, 797 (2006).


\bibitem{bode2007}  Bode, M., Heide, M., von Bergmann, K., Ferriani, P., Heinze, S., Bihlmayer, G.,  Kubetzka, A., Pietzsch, O., Bl\"ugel, S. \&  Wiesendanger, R. Chiral magnetic order at surfaces driven by inversion asymmetry. Nature {\bf447}, 190 (2007).

\bibitem{ferriani2008} Ferriani, P., von Bergmann, K., Vedmedenko, E. Y., Heinze, S., Bode, M., Heide, M., Bihlmayer, G., Bl\"ugel, S. \& Wiesendanger, R. Atomic-Scale Spin Spiral with a Unique Rotational Sense: Mn Monolayer on W(001). Phys. Rev. Lett. {\bf101}, 027201 (2008).

 \bibitem{santosCond2008} Santos, B., Puerta, J. M., Cerda,  J. I., Stumpf, R., von Bergmann, K., Wiesendanger, R., Bode, M., McCarty, K. F. \&  de la Figuera, J. Structure and magnetism of ultra-thin chromium layers on W(110). New J. Phys. {\bf10}, 013005 (2008).

 \bibitem{zimmermanPRB-2014} Zimmermann, B., Heide, M., Bihlmayer, G. \& Bl\"ugel, S. First-principles analysis of a homochiral cycloidal magnetic structure in a monolayer Cr on W(110). Phys. Rev. B  {\bf90}, 115427 (2014).

\bibitem{mentzel2012} Menzel, M., Mokrousov, Y., Wieser, R., Bickel, J. E., Vedmedenko, E., Bl\"ugel, S.,  Heinze, S., von Bergmann, K., Kubetzka, A. \& Wiesendanger, R. Information Transfer by Vector Spin Chirality in Finite Magnetic Chains. Phys. Rev. Lett. {\bf108}, 197204 (2012).

\bibitem{heinze2011} Heinze, S., von Bergmann, K., Menzel, M., Brede,  J., Kubetzka,  A., Wiesendanger, R., Bihlmayer, G. \& Bl\"ugel, S. Spontaneous atomic-scale magnetic skyrmion lattice in two dimensions. Nat. Phys. {\bf7}, 713 (2011).


\bibitem{romming} Romming, N., Hanneken, C., Menzel, M., Bickel, J. E., Wolter, B., von Bergmann,  K., Kubetzka, A. \& Wiesendanger, R. Writing and Deleting Single Magnetic Skyrmions. Science {\bf 314}, 636 (2013).

\bibitem{miron2011} Miron, I. M., Garello, K., Gaudin, G., Zermatten, P.-J., Costache, M. V., Auffret, S.,  Bandiera, S., Rodmacq, B., Schuhl, A. \& Gambardella, P. Perpendicular switching of a single ferromagnetic layer induced by in-plane current injection. Nature {\bf 476}, 189 (2011).

\bibitem{liu-science-2012} Liu, L., Pai, C.-F., Li, Y., Tseng, H. W., Ralph, D. C. \&  Buhrman, R. A. Spin-Torque Switching with the Giant Spin Hall Effect of Tantalum. Science {\bf336}, 555 (2012).

\bibitem{manchon2008}  Manchon, A. \& Zhang, S. Theory of nonequilibrium intrinsic spin torque in a single nanomagnet. Phys. Rev. B {\bf78}, 212405 (2008).

\bibitem{manchon20081} Garate, I. \&  MacDonald, A. H. Theory of nonequilibrium intrinsic spin torque in a single nanomagnet. Phys. Rev. B {\bf80}, 134403 (2009).

\bibitem{manchon20082} Haney, P. M., Lee, H.-W., Lee,  K.-J., Manchon, A. \&  Stiles, M. D. Current-induced torques and interfacial spin-orbit coupling. Phys. Rev. B {\bf88}, 214417 (2013).

\bibitem{freimuth2014} Freimuth, F., Bl\"ugel, S. \& Mokrousov, Y. Spin-orbit torques in Co/Pt(111) and Mn/W(001) magnetic bilayers from first principles. Phys. Rev. B {\bf90}, 174423 (2014). 

 \bibitem{freimuth20141} Freimuth, F., Bl\"ugel, S. \& Mokrousov, Y. Direct and inverse spin-orbit torques. Phys. Rev. B {\bf92}, 064415 (2015).

\bibitem{HeidePRB:2008} Heide, M., Bihlmayer, G. \& Bl\"ugel, S. Dzyaloshinskii-Moriya interaction accounting for the orientation of magnetic domains in ultrathin films: Fe/W(110). Phys. Rev. B  {\bf78}, 140403 (2008). 
\bibitem{thiaville2012} Thiaville, A., Rohart, S., Ju\'e, E., Cros, V. \& Fert, A. Dynamics of Dzyaloshinskii domain walls in ultrathin magnetic films. Europhys. Lett. {\bf 100}, 57002 (2012).

\bibitem{emori2013} Emori, S., Bauer, U., Ahn, S.-M., Martinez, E. \&  Beach, G. S. D. Current-driven dynamics of chiral ferromagnetic domain walls. Nat. Mater. {\bf12}, 611 (2013).


\bibitem{ryu2013} Ryu, K.-S., Thomas, L., Yang, S.-H. \& Parkin, S. Chiral spin torque at magnetic domain walls. Nat. Nanotechnol. {\bf8}, 527 (2013).

\bibitem{chen-prl-2013} Chen, G., Zhu,  J., Quesada,  A., Li,  J., N'Diaye, A. T., Huo, Y., Ma, T. P., Chen, Y., Kwon, H. Y., Won, C., Qiu, Z. Q., Schmid, A. K. \& Wu, Y. Z. Novel Chiral Magnetic Domain Wall Structure in Fe/Ni/Cu(001) Films. Phys. Rev. Lett. {\bf110}, 177204 (2013).

\bibitem{chen-natcom-2013} Chen, G., Ma, T., N'Diaye,  A. T., Kwon,  H., Won, C., Wu, Y. \&  Schmid,  A. K. Tailoring the chirality of magnetic domain walls by interface engineering. Nat. Commun. {\bf4}, 2671 (2013).

\bibitem{Tetienne2014}  Tetienne, J.-P., Hingant, T., Martinez, L. J.,  Rohart, S.,  Thiaville, A., Herrera Diez, L., Garcia, K., Adam, J.-P., Kim,  J.-V., Roch, J.-F., Miron, I. M., Gaudin, G., Vila,  L., Ocker,  B., Ravelosona,  D.,  Jacques, V. The nature of domain walls in ultrathin ferromagnets revealed by scanning nanomagnetometry. Nat. Commun. {\bf6}, 6733 (2015). 

\bibitem{hrabec2014} Hrabec, A., Porter, N. A., Wells, A., Benitez, M. J., Burnell, G., McVitie, S. ,  McGrouther, D., Moore,  T. A. \&  Marrows, C. H. Measuring and tailoring the Dzyaloshinskii-Moriya interaction in perpendicularly magnetized thin films. Phys. Rev. B {\bf 90}, 020402(R) (2014).

\bibitem{je2013}  Je, S.-G., Kim, D.-H., Yoo, S.-C., Min, B.-C., Lee, K.-J. \&  Choe, S.-B. Asymmetric magnetic domain-wall motion by the Dzyaloshinskii-Moriya interaction. Phys. Rev. B {\bf88}, 214401 (2013).

\bibitem{pizzini2014} Pizzini, S., Vogel,  J., Rohart, S., Buda-Prejbeanu, L. D.,  Ju\'e, E., Boulle, O., Miron,  I. M., Safeer, C. K., Auffret, S., Gaudin, G. \& Thiaville, A. Chirality-Induced Asymmetric Magnetic Nucleation in Pt/Co/AlOx Ultrathin Microstructures. Phys. Rev. Lett. {\bf113}, 047203 (2014).

\bibitem{Torrejon2014} Torrejon, J., Kim, J., Sinha, J., Mitani, S., Hayashi, M., Yamanouchi M. \& Ohno, H.
Interface control of the magnetic chirality in CoFeB/MgO heterostructures with heavy-metal underlayers. Nat. Commun. {\bf5}, 4655 (2014). 

\bibitem{perd-zunger-LDA} Perdew, J. P. \& Zunger, A. Self-interaction correction to density-functional approximations for many-electron systems. Phys. Rev. B {\bf23}, 5048 (1981).

\bibitem{Wimmer1981} Wimmer, E., Krakauer, H., Weinert, M. \&  Freeman, A. J. Full-potential self-consistent linearized-augmented-plane-wave method for calculating the electronic structure of molecules and surfaces: O$_{2}$ molecule. Phys. Rev. B {\bf24}, 864 (1981).

\bibitem{Fleur} URL http://www.flapw.de.


\bibitem{KurzPRB-2004} Kurz, Ph. F\"orster, F. Nordstr\"om, L. Bihlmayer, G. \& Bl\"ugel, S.
\textit{Ab initio} treatment of noncollinear magnets with the full-potential linearized augmented plane wave method. Phys. Rev. B {\bf69}, 024415 (2004).

\bibitem{Heide-phys-B} Heide, M., Bihlmayer, G. \& Bl\"ugel, S. Describing Dzyaloshinskii-Moriya spirals from first principles, Phys. B. Cond. Matt. {\bf404}, 2678 (2009).

%%\bibitem{swdft} see e.g. F. F. Schubert, Y. Mokrousov, P. Ferriani, and S. Heinze, Phys. Rev. B {\bf83}, 165442 (2011).
%%
%\bibitem{daalderop} G.H.O. Daalderop, P.J. Kelly and M.F.H. Schuurmans, Phys. Rev. B {\bf50}, 9989 (1994).

\bibitem{maca2013}  M\'aca, F., Kudrnovsk\'y,  J., Drchal,  V. \& Redinger,  J.  Influence of oxygen and hydrogen adsorption on the magnetic structure of an ultrathin iron film on an Ir(001) surface. Phys. Rev. B {\bf88}, 045423 (2013).

\bibitem{dupe}  Dup\'e, B., Hoffmann, M., Paillard, C. \& Heinze. S. Tailoring magnetic skyrmions in ultra-thin transition metal films. Nat. Commun. {\bf5}, 4030 (2014).

\bibitem{kashid} Kashid, V., Schena, T., Zimmermann, B., Mokrousov, Y., Bl\"ugel, S., Shah, V. \& Salunke, H. G. Dzyaloshinskii-Moriya interaction and chiral magnetism in $3d-5d$ zigzag chains: Tight-binding model and \textit{ab initio} calculations. Phys. Rev. B {\bf90}, 054412 (2014).

\bibitem{BornemannPRB-2012} Bornemann, S.,  \ifmmode \check{S}\else \v{S}\fi{}ipr, O., Mankovsky, S., Polesya, S., Staunton, J. B., Wurth, W., Ebert, H., \& Min\'ar, J. Trends in the magnetic properties of Fe, Co, and Ni clusters and monolayers on Ir(111), Pt(111), and Au(111). Phys. Rev. B {\bf86}, 104436 (2012).

\bibitem{kurpinPRB2005}  Krupin, O., Bihlmayer, G., Starke, K., Gorovikov, S., Prieto, J. E., D\"obrich, K., Bl\"ugel, S., and \& Kaindl, G. Rashba effect at magnetic metal surfaces. Phys. Rev. B {\bf71}, 201403(R) (2005).

\bibitem{stampflPRB} Zhang, H., Soon, A., Delley, B. \& Stampfl, C. Stability, structure, and electronic properties of chemisorbed oxygen and thin surface oxides on Ir(111). Phys. Rev. B {\bf78}, 045436 (2008).}


\end{thebibliography}
\end{document}